\documentclass[conference]{IEEEtran}
\usepackage[pdftex]{graphicx}
\usepackage{graphicx}
\usepackage{balance}
\usepackage{comment}
\usepackage{algorithm}
\usepackage[noend]{algpseudocode}
\usepackage{xcolor}
\usepackage{cite}
\usepackage{amsmath}
\usepackage{amssymb}

\DeclareMathOperator{\med}{med}

\usepackage{booktabs}

\usepackage{pgfplots}
\DeclareUnicodeCharacter{2212}{−}
\usepgfplotslibrary{groupplots,dateplot}
\usetikzlibrary{patterns,shapes.arrows}
\pgfplotsset{compat=newest}
\newlength\figH
\newlength\figW
\usepackage{lipsum}
\usepackage[capitalise]{cleveref}
\usepackage{acronym}

\acrodef{AoA}{Angle of Arrival}
\acrodef{RSSI}{Received Signal Strength Indicator}
\acrodef{SVM}{Support Vector Machine}
\acrodef{KNN}{K-Nearest Neighbors}
\acrodef{RF}{Random Forest}
\acrodef{SF}{superframe}
\acrodef{HMM}{Hidden Markov Model}
\acrodef{MLC}{machine learning classifier}
\acrodef{BLE}{Bluetooth\textsuperscript{\textregistered} Low Energy}
\acrodef{EPhESOS}{Energy and Power Efficient Synchronous Sensor Network}
\acrodef{BS}{base station}
\acrodef{MAC}{media access control}
\acrodef{TDMA}{time division multiple access}
\acrodef{ToA}{Time of Arrival}
\acrodef{TDoA}{Time Difference of Arrival}
\acrodef{SN}{sniffer node}
\acrodef{PS}{photo sensor}
\acrodef{MN}{measurement node} 
\acrodef{SAL-RB-Dataset}{SAL Autarkic Localization RSSI BLE Dataset}
\acrodef{RBF}{radial basis function}
\acrodef{IWSN}{Industrial wireless sensor network}
\acrodef{GPS}{Global Positioning System}
\acrodef{UWB}{ultra-wideband}
\acrodef{WLAN}{Wireless Local Area Network}
\acrodef{I40}{Industry 4.0}

\IEEEoverridecommandlockouts

\begin{document}
\title{
RSSI-Based Machine Learning with Pre- and Post-Processing for Cell-Localization in IWSNs

}

 \author{\IEEEauthorblockN{Julian Karoliny\IEEEauthorrefmark{1}, Thomas Blazek\IEEEauthorrefmark{1}, Fjolla Ademaj\IEEEauthorrefmark{1}, Hans-Peter Bernhard\IEEEauthorrefmark{1}\IEEEauthorrefmark{2}, Andreas Springer\IEEEauthorrefmark{2}\\
 \IEEEauthorblockA{\IEEEauthorrefmark{1}Silicon Austria Labs GmbH, 4040 Linz\\\IEEEauthorrefmark{2}Johannes Kepler University Linz, Institute for Communications Engineering and RF-Systems, 4040 Linz, Austria}}\thanks{This work is funded by the InSecTT project (https://www.insectt.eu/). InSecTT has received funding from the ECSEL Joint Undertaking (JU) under grant agreement No 876038. The JU receives support from the European Union’s Horizon 2020 research and innovation programme and Austria, Sweden, Spain, Italy, France, Portugal, Ireland, Finland, Slovenia, Poland, Netherlands, Turkey. The document reflects only the author’s view and the Commission is not responsible for any use that may be made of the information it contains.}
 }

\maketitle

\begin{abstract}
Industrial wireless sensor networks are becoming crucial for modern manufacturing. If the sensors in those networks are mobile, the position information, besides the sensor data itself, can be of high relevance. E.g. this position information can increase the trustability of a wireless sensor measurement by assuring that the sensor is not physically removed, off track, or otherwise compromised.

In certain applications, localization information at cell-level, whether the sensor is inside or outside a room or cell, is sufficient. For this, localization using Received Signal Strength Indicator (RSSI) measurements is very popular since RSSI values are available in almost all existing technologies and no direct interaction with the mobile sensor node and its communication in the network is needed. For this scenario, we propose methods to improve the robustness and accuracy of common machine learning classifiers, by using features based on short-term moments and a second classification stage using Hidden Markov Models. With the data from an extensive measurement campaign, we show the applicability of our method and achieve a cell-level localization accuracy of 93.5\%.
\end{abstract}
\vspace{0.2cm}

\section{Introduction}
\label{sec:introduction}
In industrial environments, sensors are traditionally connected through a wired communication network like field buses or Ethernet networks. However, wireless communication is becoming crucial to advanced manufacturing~\cite{IWSN_survey} and acts as an enabler for Industry 4.0. \acp{IWSN} must meet stringent reliability and latency requirements \cite{Montgomery2020a}, but offer advantages like mobile operation, easy sensor replacement, flexible mounting, and often lower cost~\cite{IWSN_Raza}. For many use cases, it is necessary to record the spatial position of the wireless sensor in addition to its measured value. As an example, we present an extension to an \ac{IWSN}-based measurement system~\cite{bernhard2020a} for the emission certification of cars according to the Euro 6 standard which traces the required measurements in time and position \cite{european2012commission}. During these tests, cars are moved between differently conditioned areas and for the position tracking a non-interfering add-on-localization extends the wireless measurement system. Besides the information about the location itself, the position information of sensor nodes is used to verify the measurements, e.g. to assure that a sensor is not physically removed, off track, or otherwise compromised. For instance, a malicious sensor node from an unverified location can be identified and its measurement values are rejected.

As the use of \ac{GPS} is strongly limited in indoor environments, factory communication systems have to use alternative  localization systems. In \acp{IWSN}, the main techniques for localization are based on \ac{AoA}, \ac{ToA}, \ac{TDoA} and \ac{RSSI}~\cite{localization_survey}. Localization based on \ac{RSSI} values is one of the most promising solutions for low-cost applications since the \ac{RSSI} value is available in existing technologies like \ac{BLE}, \ac{WLAN}, ZigBee, etc. However, due to multipath fading, noise and limited dynamic range of the \ac{RSSI} measurements, exact localization based on a path-loss model and multilateration becomes quite challenging. While in the literature many techniques focus on improving the accuracy of \ac{RSSI}-based estimation, there are also many use cases in \acp{IWSN}, where a coarse location of the sensor node is sufficient, such as measurement verification, security, and automotive testing. The main task in such use cases is to classify specific environments or regions like  the inside or outside of a room and determine whether a sensor node belongs to such a confined region. Authors in \cite{indoor_cell_level_localization_based_on_RSSI_classification,RSSI_based_classification_for_indoor} have studied the so-called cell-level-based localization with \ac{RSSI} values using supervised machine learning methods. A major challenge with these methods is the limited amount of training and validation data.
\subsection{Contribution}
In this work, we present a \ac{RSSI}-based cell-level localization approach as an add-on to an existing \ac{IWSN}.
To acquire the \ac{RSSI} measurements of all sensor values sent to the \ac{BS}, we use additional sensor nodes which only listen passively. 
We  propose  methods  to  improve  the robustness and accuracy of common \acp{MLC}, by  introducing suitable input features  and  a  subsequent second classification stage.
Here, we take advantage of the fact that the \ac{RSSI} measurements are highly correlated in time, i.e., two subsequent measurements are from similar positions because of the limited movement speed. Additionally, we conducted an extensive measurement campaign that allows us to test and verify the localization method we developed.
\subsection{Notation}
Scalars are written as $x$, while vectors and matrices are denoted as lower- and uppercase bold respectively ($\mathbf{x}$ and $\mathbf{X}$). For matrices and vectors, the element-wise (Hadamard) multiplication is denoted with $\circ$.
Time indices are indicated with superscript $x_t$ and a set of time depending measurements of the size $T$ is written as $x_{1:T} \equiv \left[ x_1, x_2, \ldots x_T \right]$.  
\section{Experimental Setup and Data Acquisition}
\label{sec:measurement_setup}
\begin{figure}[ht]
  \centering
    \includegraphics[width=0.5\textwidth]{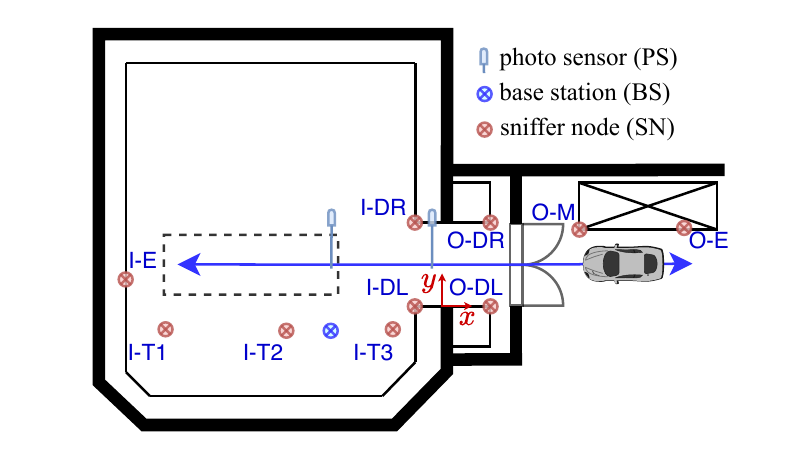}
    \vspace{-0.5cm}
    \caption{Measurement setup. The Car is driving in and out of the automotive testbed.}
    \label{fig:measurement_setup}
\end{figure}

All measurements in this work were obtained at an automotive testbed\footnote{Chassis dynamo-meter, AVL List GmbH in Graz, Austria}, i.e. under real-world conditions. In our measurement scenario, we want to estimate the  position of a wireless sensor node mounted on a car that is moving in and out from the testbed, c.f.~\cref{fig:measurement_setup}. This sensor node is referred to  as \ac{MN}. Additionally, ten \acp{SN} using the same transceiver as the \ac{MN} are placed inside and outside the testbed, referred to as I-x respectively O-x. The \acp{SN} acquire the \ac{RSSI} values of every communication packet sent from the \ac{MN} to the \ac{BS}. Additionally, two \acp{PS} are placed at the doorstep and the car test-position respectively, to automatically label the measurements. The \acp{PS} are only used during the measurement campaign for the labeling task. For the operation phase the \acp{PS} would not be sufficient for the localization task, as the application requires to localize and distinguish multiple cars in different clusters.

\Cref{fig:measurement_setup} depicts the measurement setup with the position of the \ac{SN}, the \acp{PS}, and the trajectory of the car. The labels and coordinates of each \ac{SN} are listed in \cref{table:pos_sniffer_nodes}.
\begin{table}
\centering
\caption{Label and position of sniffer-nodes}
\label{table:pos_sniffer_nodes}
\begin{tabular}{@{}ll|rrr@{}}
\toprule
      &                    & \multicolumn{3}{c}{Position in m} \\ 
Label & Description        & x          & y         & z        \\ \midrule
I-E   & inside end         & -13.30     & 0.90      & 0.64     \\
I-T1  & inside top 1       & -9.88      & -0.35     & 3.92     \\
I-T2  & inside top 2       & -6.58      & -0.35     & 3.92     \\
I-T3  & inside top 3       & -1.60      & -0.35     & 3.92     \\
I-DR  & inside door right  & -1.55      & 3.19      & 1.00     \\
I-DL  & inside door left   & -1.55       & -0.09     & 1.00     \\ \midrule
O-E   & outside end        & 11.10      & 2.02      & 1.70     \\
O-M   & outside mid        & 6.80        & 2.32     & 1.84     \\
O-DR  & outside door right & 2.21       & 3.10      & 0.99     \\
O-DL  & outside door left  & 2.21       & 0.00      & 1.00     \\ \bottomrule
\end{tabular}
\end{table}
\subsection{Hardware and Protocol}
We used the \ac{BLE} physical (PHY) layer and combined it with the \ac{EPhESOS} protocol~\cite{life_cycle_ephesos} to realize an \ac{IWSN} with up to 100 nodes per \ac{BS}. \ac{EPhESOS} provides a deterministic \ac{MAC} layer using \ac{TDMA}, with a \ac{SF} length of 100\,ms. As hardware platform for the measurements and all applications, the Nordic\texttrademark NRF52840 controller with integrated transceiver is used. 

\subsection{Acquired Dataset}
\begin{figure}[t]
  \centering
    \includegraphics[width=0.5\textwidth]{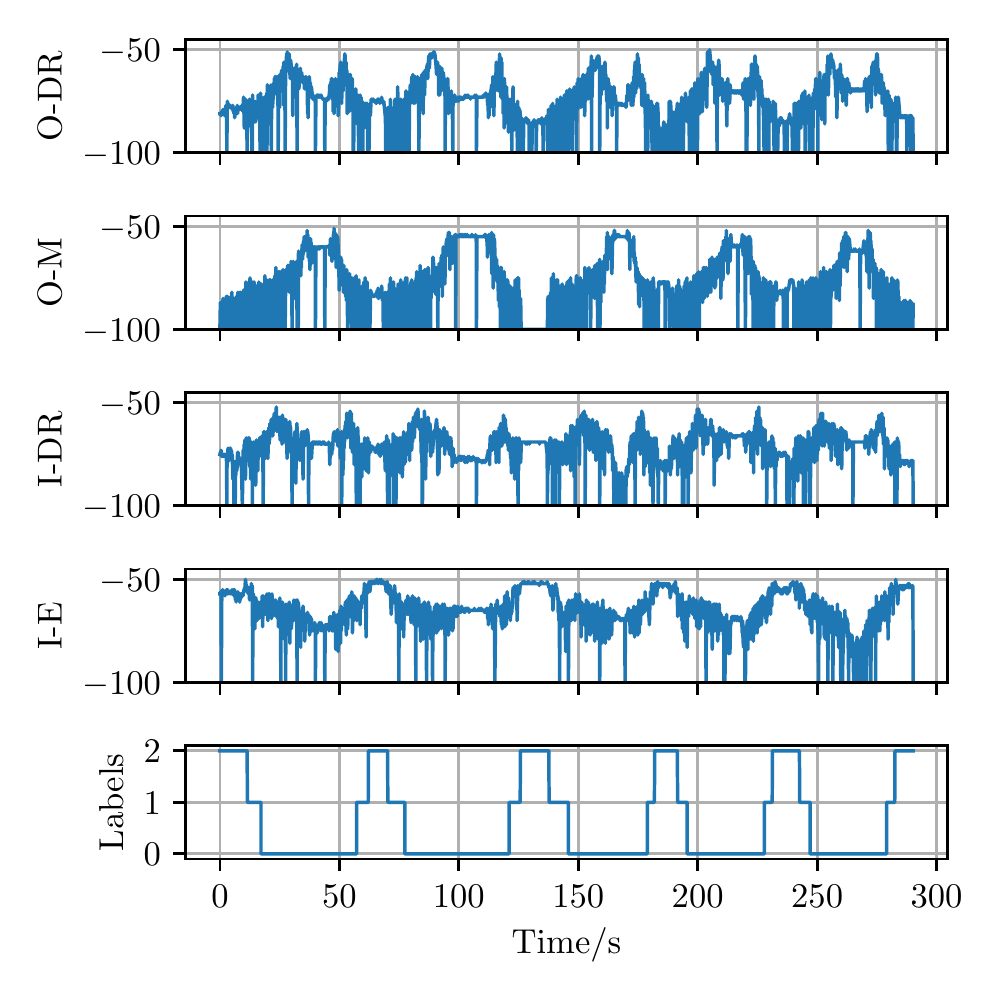}
    \vspace{-1cm}
    \caption{Exemplary RSSI measurements in dBm of four SNs with the corresponding labels.}
    \label{fig:example_rssi_measurement}
\end{figure}

In the course of this work, a large number of measurements were collected, which are published and provided  as open-source data set under \ac{SAL-RB-Dataset} \cite{SAL-RB_Dataset}. The acquired data set consists of: (i) two disjoint measurement-sets, where a person walks inside and outside the automotive testbed, and (ii) eight disjoint measurement-sets, where a car is driving in and out the testbed as depicted in \cref{fig:measurement_setup}. For (ii) two different \ac{MN} were mounted on the car to investigate the effects of different hardware on the classification. Each individual set provides \ac{RSSI} values of all ten \ac{SN} in a 100\,ms interval. A missing link is denoted with -100\,dBm.
The labels of the \ac{RSSI} values correspond to different localization cells and are defined as follows:
\begin{itemize}
    \item Label = 0: \,car is outside the testbed
    \item Label = 1: \,car is completely inside the testbed
    \item Label = 2: \,car is inside the testbed and on test-position
\end{itemize}
In total the data set consists of more than 20\,000 labeled \ac{RSSI} measurements for each \ac{SN}. \Cref{fig:example_rssi_measurement} shows example measurements of the \ac{RSSI} values in dBm for the \acp{SN} $\lbrace\text{I-E, I-DR, O-M, O-DR}\rbrace$ and the corresponding labels.
To evaluate the findings of this work, six car measurement-sets from the \ac{SAL-RB-Dataset} are used.
\section{RSSI-based Machine Learning Classifier}
\label{sec:model}
The \ac{MLC} uses the sniffed \ac{RSSI} values of data packets sent by the \ac{MN} to identify its position, or more precisely the label of the corresponding cell in the testbed.
The \ac{RSSI} measurements from the ten \acp{SN} ($N=10$) are synchronously recorded at time step $t$ and collected in the input vector $\mathbf{x}_t \in \mathbb{R}^N$ which is used to estimate the corresponding label $y_t \in \lbrace0, 1, 2\rbrace$. As usual, the \ac{MLC} has an offline training phase and an online classifier phase. The training set $\lbrace \mathbf{x}^{\prime}_{1:T^{\prime}},\,y^{\prime}_{1:T^{\prime}} \rbrace$ with $T^{\prime}$ samples is used to train the classifier. Afterwards the classification is performed online with the measured data to estimate  the label $\hat{y}_t$. The performance of the classification is assessed by the accuracy score which is calculated as the number of correctly identified labels divided by the number of all classifications.

\subsection{Machine Learning Methods and Data Splitting}
In the following, we focus on simple machine learning techniques to enable the estimator implementation directly at node level. We considered three widely used algorithms, namely \ac{KNN}, \ac{RF}, and \ac{SVM}. All of these algorithms fall into the category of supervised machine learning techniques, where the choice of measurement-sets for the training and subsequent validation is crucial. A common procedure here is to perform a random split of the measurement data to obtain a subset for training and validation. However, since the \ac{RSSI} values are measured continuously, subsequent measurements tend to be very similar. A random split would lead in this case to a very good, though, unrealistic accuracy score, as the smaller validation set contains nearly identical measurements of the training set. To avoid this, we do not split or shuffle individual measurements set, but keep them whole either for training or testing. This approach is also comparable to the real use-case since here the \ac{MLC} would also be learned at the beginning and should then work for future measurements.
Additionally, instead of using only a single training and validation set, all combinations consisting of three training and one validation data sets, without using the same for both, are evaluated in this section. This ensures a fair comparison of the different \ac{MLC} since it mitigates the problem that some approaches may be exceptionally good for some data set combinations. 

The three proposed algorithms are implemented using Scikit-learn which is an open-source machine learning library for Python \cite{scikit-learn}. For the given classification task, all three algorithms performed similarly in terms of accuracy and robustness, though the \ac{SVM} showed slightly higher accuracy. With the \ac{RSSI} values directly as input, the \ac{SVM} reached an average accuracy of 77\,\% for the given data set \cite{SAL-RB_Dataset} and is used exclusively for all following evaluations.

The remaining miss-classifications are caused by overlapping class-conditional distributions due to noise and limited dynamic range of the measurements. These  mainly occur at class transition regions in the testbed, e.g., the doorstep and test-position, and none of the proposed \ac{MLC} could improve in these areas. In order to further improve the accuracy, especially at the class boarders, we introduce more suitable input features for the \ac{MLC} and a subsequent post-processing.  

\subsection{Selecting Features for Machine Learning Classifier}
\label{sec:feature_selection}
Regarding feature selection, two questions have to be answered. Firstly, it is necessary to know whether the raw data $\mathbf{x}_t$ of one \ac{SF} is sufficient for the classification, or if including previous samples will improve the performance. Secondly, it is essential to validate how the position, number, and combination of the \acp{SN} influences the accuracy. In this context, it is important to answer if also a subset of nodes is sufficient for the classification task. 

Also for this evaluation, the choice of measurement-sets for the training and subsequent validation is crucial. To ensure a fair comparison of the different features all combinations of three training and one validation data sets, without using the same for both, are evaluated. This mitigates the problem that some features may be exceptionally good for some data set combinations

\vspace{0.05cm}\subsubsection{Short-Term Moments as Features}
\label{subsec:short_term_moments}
The drawback of using the \ac{RSSI} measurements of more than one \ac{SF} as input features is, that it increases the feature space with each additional sample. Besides that, the selected \ac{MLC} may not be able to model the relation of sequential input data, e.g. \ac{RSSI} values are considered individually and the dynamic relation over time is not modeled. 

To avoid this, we propose to use short-term estimates of the first two moments over \(L\) samples. Thus, we introduce 
\begin{equation}
    \mathbf{x}_t^{\mu \sigma} = \Bigg[\underbrace{\frac{1}{L}\sum_{\tau=t-L+1}^t{x}_{\tau}}_{\bar{x}^L_t}, \quad \underbrace{\frac{1}{L-1}\sum_{\tau=t-L+1}^t({x}_{\tau}-\bar{x}^L_t)^2}_{s_t^L}\Bigg] \,.\label{eq:mu-sigma}
\end{equation}
This only doubles the feature space, independently of the number of used \acp{SF}. To get an indication of how many preceding samples benefit this approach in our scenario, we consider the coherence time of the channel. The coherence time $T_c$ is a statistical measure of the time duration over which two received signals have a certain minimum  amplitude correlation, depending on the relative motion between the \ac{MN} on the car and the \acp{SN}. It can be estimated according to \cite{rappaport1996wireless} with
\begin{equation}
    T_c = \sqrt{\frac{9}{16\pi f_m^2}} = \frac{0.423}{f_m} \,, \label{eq:coherence_time}
\end{equation}
where $f_m$ is the  Doppler spread, which is upper bounded by the maximum occurring Doppler shift, calculated as $f_{max} = v / \lambda $. With the given \ac{BLE} center frequency $f_c = 2.44\,\text{GHz}$ and the average speed of the car $v = 1\,\frac{\text{m}}{\text{s}}$, the Doppler shift in case of directly oncoming movement is about 8\,Hz, which results in a lower bound of $T_c \approx 50\,\text{ms}$. Since the time interval between two successive \acp{SF} is 100\,ms, considering more than two \ac{RSSI} measurements for (\ref{eq:mu-sigma}) may not improve the result.

\begin{figure}
  \centering
    \includegraphics[width=0.5\textwidth]{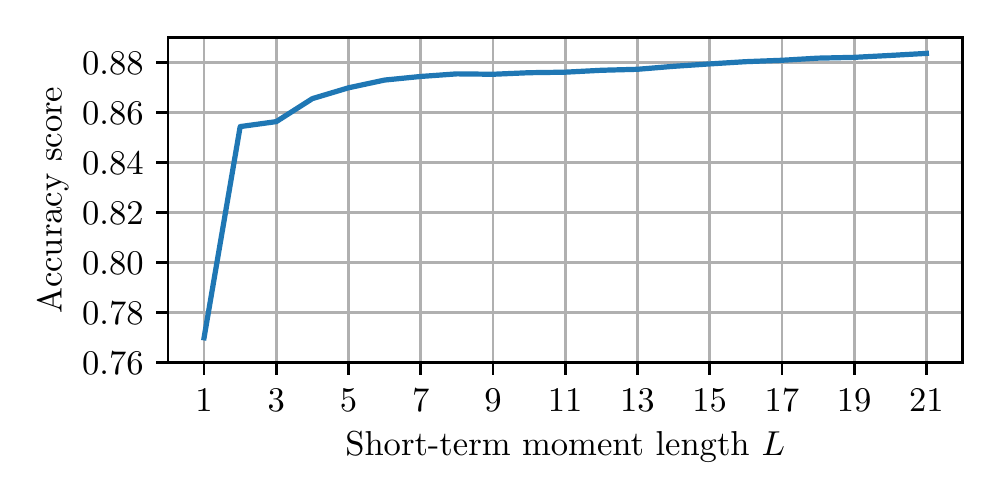}
    \vspace{-1cm}
    \caption{Accuracy score over increasing short-term moment length $L$ for the \ac{SVM}. The raw \ac{RSSI} measurements are indicated with $L=1$.}
    \label{fig:acc_vs_musigma_len}
\end{figure}

To investigate this and to analyze the advantages of using short-term moments as input features, we use the \ac{SVM} classifier  and compare the results for an increasing number $L$ in (\ref{eq:mu-sigma}). \Cref{fig:acc_vs_musigma_len} depicts the accuracy score of the classifier over $L$, where $L=1$ denotes the result using the raw \ac{RSSI} values. The depicted accuracy is the average accuracy over all $1023$ possible \ac{SN}-combinations and 60 possible training and validation data set combinations. As mentioned before, this assures a fair comparison since we observed that the short-term moments showed a higher improvement for certain combinations. In contrast to the calculated coherence time, the accuracy  further increases with $L$, though the highest relative gain is achieved by using one additional preceding measurement. 

\vspace{0.15cm}\subsubsection{Node Selection Scheme}
Due to noise and the limited dynamic range of the measurements, two \acp{SN} can provide similar \ac{RSSI} values, though they are at different locations. 
Additionally, some \ac{SN} positions may provide exceptional good measurements for the given classification task, since the main effect that is exploited is not the free-space pathloss, but the changes between line-of-sight and non-line-of-sight channel caused by the movement of the car. Not only the positions of the \acp{SN} are important, but also the combination of the individual \ac{RSSI} measurements is crucial for the correct classification. Instead of evaluating all \ac{SN} combinations and simply choosing the one with the highest accuracy, we first determine combinations that provide poor performance.

\begin{figure}
  \centering

    \includegraphics[width=0.5\textwidth]{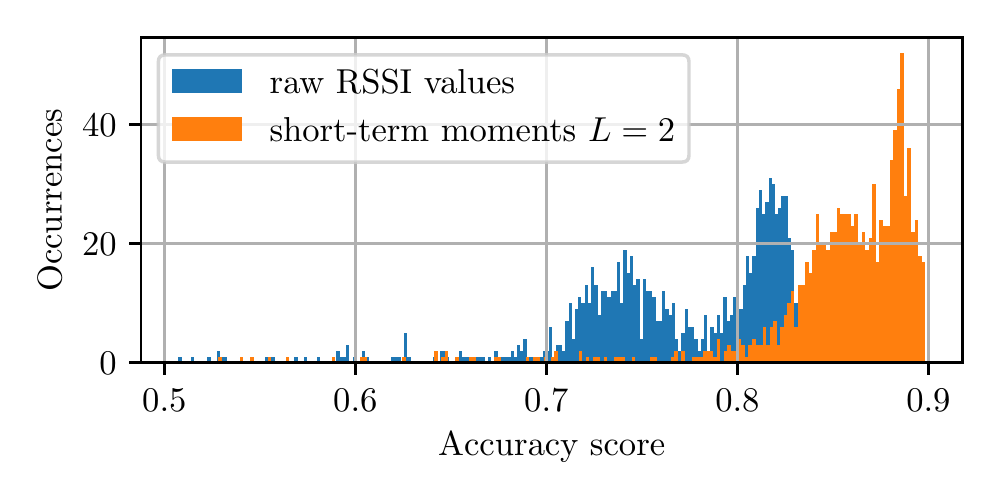}
    \vspace{-1cm}
    \caption{Accuracy score distribution of all \ac{SN} combinations using the raw \ac{RSSI} measurements and short-term moments for the \ac{SVM}.}
    \label{fig:hist_sensor_combinations}
\end{figure}

\Cref{fig:hist_sensor_combinations} depicts the distribution of the accuracy score of all 1023 \ac{SN}-combinations averaged over all data set combinations using the raw \ac{RSSI} values and the short-term moments in (\ref{eq:mu-sigma}) with $L=2$. Again the advantages of the short-term moments can be observed, as all individual \ac{SN}-combinations show a higher accuracy. Additionally, they are clustered at higher percentages which indicate improved robustness.
Because the short-term moments are superior as input features, in the following we will use them exclusively.

Most \ac{SN}-combinations show an accuracy between 80-90\,\%, tending to the higher ones, while only very few combinations are below. On closer inspection of the few combinations that lead to poor results, we found an intuitive explanation. These combinations are composed of either \acp{SN} only outside, only inside, or only single \acp{SN}. On the contrary, combinations that are composed of \acp{SN} equally spaced in the area of interest, including nodes at significant points, e.g. near the doorstep, lead to very good results. We found out that about four \acp{SN} are sufficient for our task, with for example the combination $\lbrace\text{I-E, I-DR, O-M, O-DR}  \rbrace$.
\subsection{Refining Cell Estimates Via Post-Processing}
The physical cell-boarders in the $x$-dimension, as depicted in \cref{fig:measurement_setup} with the \acp{PS}, are defined w.r.t. the given localization task and are not chosen optimal in terms of high differences in the measured \ac{RSSI} values. As a result, noise and limited dynamic range of the \ac{RSSI} measurements lead to oscillating miss-classifications in time, especially near the physical borders of the cells, i.e. doorstep and test-position. Miss-classifications can also occur in the middle of the cell, e.g. inside the testbed at the far end, which in particular is a problem for the mentioned industrial use-case.

Therefore, we propose a second classification stage to mitigate this problem.
Instead of considering \acp{SF} individually, we include a certain dynamic in the classification model, because the samples are highly correlated in time. For example preceding measurements have a high probability to result in the same cell and abrupt changes for a few \acp{SF} are physically not possible.
The two stage-approach consist of, (i) the \ac{MLC} for the current input vector $\mathbf{x}_t$ or $\mathbf{x}_t^{\mu \sigma}$ respectively and the corresponding output $\hat{y}_t$, and (ii) a filter to account for the dynamic in the cell transitions with the output $\hat{z}_t$.
For (ii) we propose two approaches, a simple median filter and a filter based on \acp{HMM} \cite{TimeSeriesClassification_HiddenMarkov}. 
\subsubsection{Median Filter}
By applying a median filter we mitigate the abrupt changes in the cell estimate. The output of the \ac{MLC} $\hat{y}_t$ at the time step $t$ is filtered using $M$ past predictions with a windowed median filter
\begin{equation}
	\hat{z}_t =  \med\big(\hat{y}_{t-M},\dots ,\hat{y}_{t}\big) \,, \label{eq:median}
\end{equation}
where $\hat{y}_t \in \lbrace 0,1,2 \rbrace$. 
The median filter is only able to improve the classification if the first stage already provides a sufficiently good result with only a few errors. It also does not account for the probabilities of the individual cell transitions or whether these cell transitions are even possible, e.g. a change from $0$ to $2$ and vice versa is not possible in our scenario.
\vspace{0.15cm}\subsubsection{Hidden Markov Model}
Here, we assume that the observed cell estimates are corrupted versions of the true cell positions and impose probabilities both for the cell transitions, as well as for corrupt observations. This allows us to define a model for the cell transitions instead of the purely empirical median filter approach. To limit the complexity of the post-processing, we assume that the Markov property is satisfied, that is, the cell or label at time $t+1$ is conditionally independent on the past, given the current cell estimate at time $t$, or
\begin{equation}
p\big(z_{t+1}|z_{t},z_{t-1}, \dots \big)  = p\big(z_{t+1}|z_{t}\big) \,,
\end{equation}
with $z_t\in\{S^1, S^2, \dots, S^n\}$. $S^i$ refers to the $i$th possible true cell location. In our case, we observe the original cells, but with possible flips between true and observed cells, hence $y_t\in\{S^1, S^2, \dots, S^n\}$ holds as well.
We use a \ac{HMM} with $n=3$ hidden states which is defined by
\begin{itemize}
    \item the transition matrix $\mathbf{A} \in \mathbb{R}^{n\times n}$  with the probabilities $a_{ij} = p\big(z_{t+1} = S^j|z_{t} = S^i\big) $ of transition from state $S^{i}$ to state $S^{j}$,
    \item the emission matrix $\mathbf{B} \in \mathbb{R}^{n\times n}$ with the probabilities $b_{ij} = p\big(\hat{y}_t = S^j|z_{t} = S^i\big) $ to observe $\hat{y}_t = S^j$ in the state $S^i$, and
    \item the state probability vector $\boldsymbol{\pi}_t \in \mathbb{R}^{n}$ \ with the probabilities $\pi_{t,i} = p\big(z_{t} = S^i\big)$ that $S^i$ is the cell location at time $t$. $\boldsymbol{\pi}_0$ denotes the initial state.
\end{itemize}
The \ac{HMM} takes the sequence of estimates $\hat{y}_t$ of the \ac{MLC} as observations and returns a sequence of cell estimates $\hat z_t$ as output.
The finding of suitable parameters $\mathbf{A}$, $\mathbf{B}$ and $\boldsymbol{\pi}_0$ is also referred to as learning problem. 
Since the \ac{HMM} should correct the miss-classified samples, an intuitive approach is to use the mistakes from the \ac{MLC} in the training phase, i.e. calculate the transition and emission matrix with the training data $\lbrace \mathbf{x}^{\prime}_{1:T^{\prime}},y^{\prime}_{1:T^{\prime}} \rbrace$  and the corresponding prediction $\hat{y}^{\prime}_{1:T^{\prime}}$.
The transition matrix $\mathbf{A}$ describes the probability of each state transition, hence, we estimate the entries with the labels of the training data $y^{\prime}_{1:T^{\prime}}$ according to
\begin{equation}
\hat{a}_{i j}=\frac{\sum_{t=1}^{T^{\prime}-1} \delta[y^{\prime}_t,S^i]\delta[y^{\prime}_{t+1},S^j]}{\sum_{t=1}^{T^{\prime}} \delta[y^{\prime}_t,S^i]}\,,
\end{equation}
where $\delta[a,b]$ is the Kronecker delta function which equals 1 for $a=b$ and 0 otherwise.
For the estimation of the emission matrix $\mathbf{B}$, we use the confusion matrix $\mathbf{C}$ of the predicted labels $\hat{y}^{\prime}_{1:T^{\prime}}$, where the entry $c_{ij}$ represents the number of samples where the original label of the training data is $S^i$, but $S^j$ was predicted. Since the entries of the emission matrix define the probabilities of a cell depending on an observation, we can directly use the normalized confusion matrix for the estimation with
\begin{equation}
\hat{b}_{i j}=\frac{c_{j i}}{\sum_{i=1}^{n} c_{j i}}\,.
\end{equation}
For the choice of the initial vector $\boldsymbol{\pi}_0$ we have two options assuming the initial state is unknown: (i) calculate the probability of each cell by counting the occurrences in the training data $y^{\prime}_{i:T^{\prime}}$ or (ii) assume an uninformative prior, where each state has the same probability. In this work, the more general case (ii) is chosen and the entries of the initial vector are estimated by
\begin{equation}
\hat{\pi}_{0,i} = \frac{1}{n} \,,
\end{equation}
where $n$ is again the number of states.

In the so-called decoding problem, the learned \ac{HMM} is used to find the most likely state sequence $\hat{z}_{1:T}$ of the model that produced the observation $\hat{y}_{1:T}$. This problem is usually solved using the Viterbi algorithm with the drawback that it needs a sequence of observations for the prediction. In this work, a simple forward approach is used to perform cell predictions in online fashion using the \emph{forward algorithm}. 
The forward algorithm calculates the state probabilities $\boldsymbol{\pi}_t$ at a certain time step $t$ using the previous state probability and the current observation. The algorithm is initialized with $\boldsymbol{\pi}_{1} = \boldsymbol{\pi}_{0} \circ \left(\mathbf{e}_{y_1}\mathbf{B}\right)$, where $\mathbf{e}_n$ is the canonical basis row vector with $1$ in the $n$th element and $0$ otherwise. For each following observation the state probability is calculated with
\begin{equation}
\boldsymbol{\pi}_{t} = \left(\frac{\boldsymbol{\pi}_{t-1}\mathbf{A}}{\sum_{i=1}^{n}\pi_{t-1,i}}  \right) \circ  \left(\mathbf{e}_{y_t}\mathbf{B}\right) \,. \label{eq:hmm_forward}
\end{equation} 
The estimation of the cell $\hat{z}_t$ is given by the state with highest probability in $\boldsymbol{\pi}_{t}$.
\section{Classifier Evaluation}
\begin{figure}[t]
  \centering
    \includegraphics[width=0.5\textwidth]{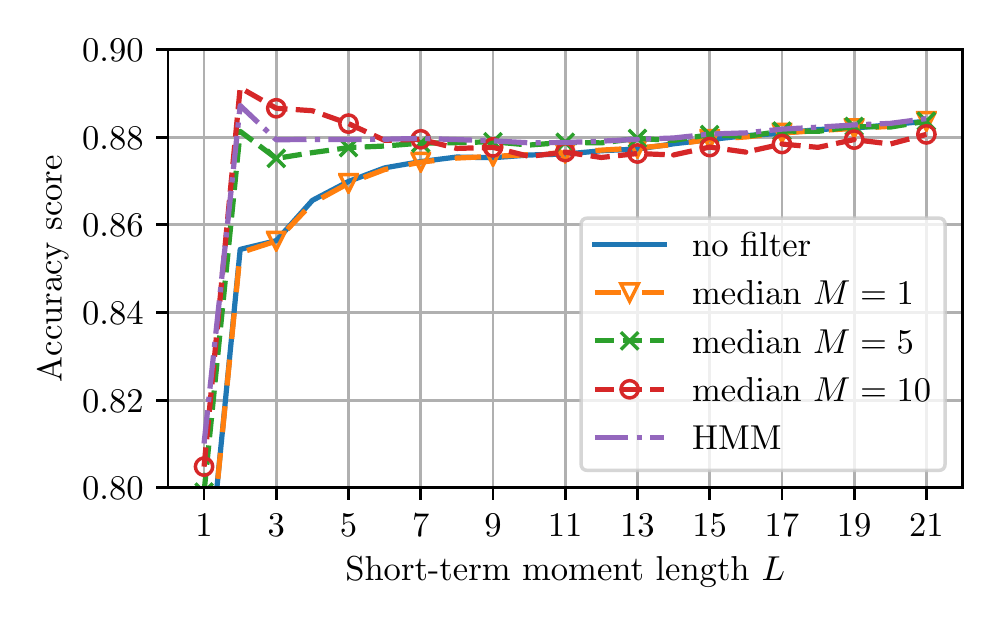}
    \vspace{-1cm}
    \caption{Accuracy score over increasing short-term moment length $L$ with and without additional filtering for the \ac{SVM}}
    \label{fig:full_acc_vs_musigma_len}
\end{figure}
Both proposed post-processing approaches are compared using the average accuracy score similar to \cref{subsec:short_term_moments} and the results are depicted in \cref{fig:full_acc_vs_musigma_len} for increasing $L$ in (\ref{eq:mu-sigma}).
For the median filter three different window lengths $M = \lbrace1, 5, 10\rbrace$ are considered. It can be observed that both approaches are able to improve the accuracy of the classification. In contrast to the raw output of the \ac{MLC}, the accuracy does not improve with increasing $L$ after the proposed filtering. The highest accuracy is achieved in both cases for $L=2$, which matches the fact that more than $2$ samples significantly oversteps our calculated coherence time estimate.

Although the median filter shows slightly better results for a sufficient window length, the \ac{HMM} is favourable since it can be adapted easily to various other classification problems and for (\ref{eq:hmm_forward}) only the state probability of the preceding sample is needed. The median filter requires storing more previous predictions and  the accuracy does not necessarily improve with increasing  window lengths. It only smooths the output of the classifier by preventing abrupt cell changes, however, this might not be suitable for all applications.

\begin{figure}[t]
  \centering
   \begin{minipage}[b]{1.0\linewidth}
    \centering
    \centerline{\includegraphics[width=1.0\linewidth]{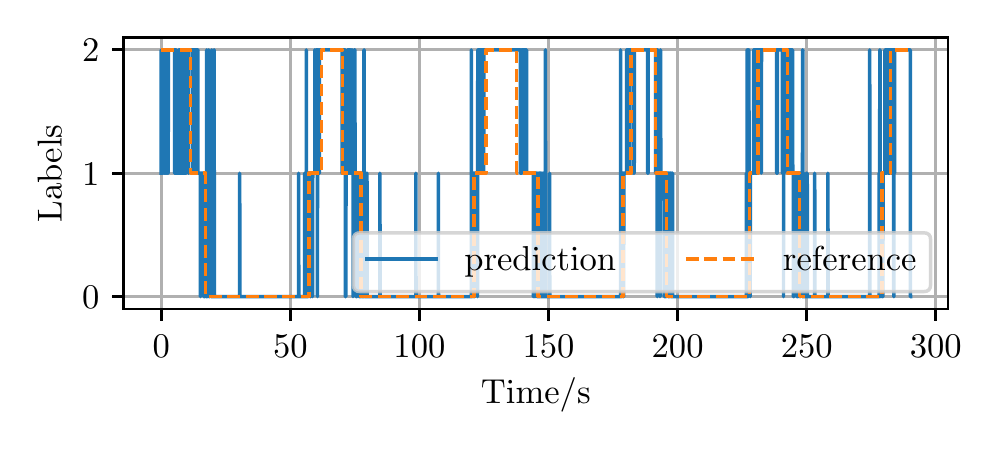}}
  \vspace{-0.2cm}
    \centerline{\small (a) raw RSSI measurements}\smallskip
  \end{minipage}
  \begin{minipage}[b]{1.0\linewidth}
    \centering
    \centerline{\includegraphics[width=1.0\linewidth]{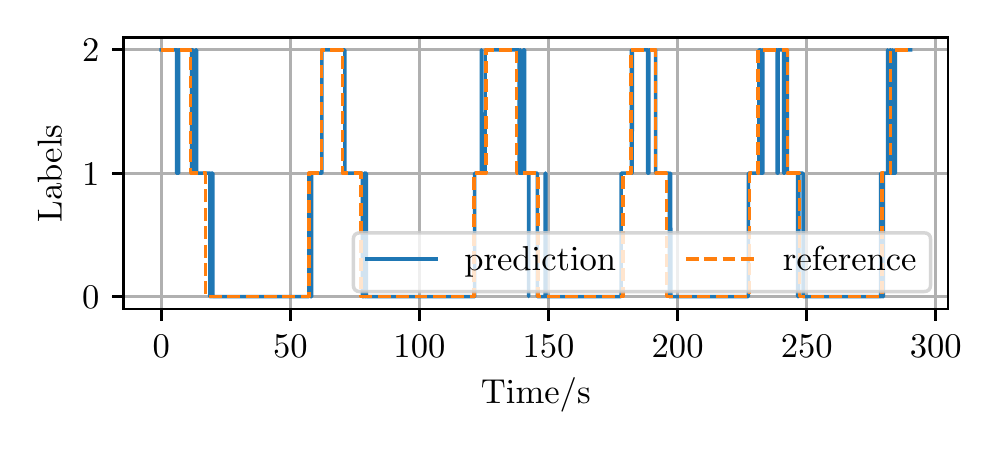}}
    \vspace{-0.2cm}
    \centerline{\small (b) short-term moments $L =2$ and HMM}
  \end{minipage}
\caption{Predicted labels compared to ground truth reference using the  \ac{SN} combination $\lbrace\text{I-E, I-DR, O-M, O-DR}\rbrace$ using the \ac{SVM} as \ac{MLC}} \label{fig:prediction_example}
\end{figure}

In the following, the classification is performed with the \ac{SN} combination $\lbrace\text{I-E, I-DR, O-M, O-DR}  \rbrace$ and a single data set combination. \Cref{fig:prediction_example} depicts the results of the \ac{MLC} for: (a) the raw \ac{RSSI} measurements with an accuracy of 86.2\,\%, and (b) the short-term moments with $L=2$ and an additional \ac{HMM} filtering with an accuracy of 93.5\,\%. Note that in contrast to \cref{fig:acc_vs_musigma_len}, the accuracy is higher since here an adequate \ac{SN} combination was chosen.

\balance
\section{Conclusion}
We analyzed the performance of cell-level localization based on \ac{RSSI} values, measured in an already existing \ac{IWSN}. The evaluation showed that for this task the accuracy of the used \ac{MLC} was comparable to each other, while the choice of good data pre- and post-processing was the key to higher accuracy. The introduced features based on short-term moments significantly increased the accuracy and robustness of the \ac{MLC} by considering only one preceding \ac{RSSI} measurement. To mitigate abrupt changes of the estimate at the output of the \ac{MLC} we added an additional classification stage. Here the \ac{HMM} showed excellent cell-level localization results with an accuracy of $93.5\,\%$. Due to the learning based on the confusion matrix and training data, it  can be adapted to various other classification problems. Based on an extensive measurement campaign we were able to test the algorithms in detail and also investigate the importance of \ac{SN} position and combination.
The proposed approach can be easily implemented at node-level to directly label the measurements or verify them based on their location.

\bibliographystyle{IEEEbib}
\bibliography{ZoneDiscretizer}

\begin{thebibliography}{10}

\bibitem{IWSN_survey}
A.~A. {Kumar S.}, K.~{Ovsthus}, and L.~M. {Kristensen.},
\newblock ``{An Industrial Perspective on Wireless Sensor Networks — A Survey
  of Requirements, Protocols, and Challenges},''
\newblock {\em IEEE Communications Surveys Tutorials}, vol. 16, no. 3, pp.
  1391--1412, 2014.

\bibitem{Montgomery2020a}
K.~Montgomery, R.~Candell, Y.~Liu, and M.~Hany,
\newblock ``Wireless user requirements for the factory workcell,''
\newblock Tech. {R}ep., National Institute of Standards and Technology, jan
  2020.

\bibitem{IWSN_Raza}
S.~Raza, M.~Faheem, and M.~Guenes,
\newblock ``{Industrial wireless sensor and actuator networks in industry 4.0:
  Exploring requirements, protocols, and challenges—A MAC survey},''
\newblock {\em International Journal of Communication Systems}, vol. 32, no.
  15, pp. e4074, 2019.

\bibitem{bernhard2020a}
H.-P. {Bernhard}, J.~{Karoliny}, B.~{Etzlinger}, and A.~{Springer},
\newblock ``Work-in-progress: Rssi-based presence detection in industrial
  wireless sensor networks,''
\newblock in {\em 2020 16th IEEE International Conference on Factory
  Communication Systems (WFCS)}, 2020, pp. 1--4.

\bibitem{european2012commission}
European Commission.,
\newblock ``Commission regulation (eu) no 459/2012 of 29 may 2012 amending
  regulation (ec) no 715/2007 of the european parliament and of the council and
  commission regulation (ec) no 692/2008 as regards emissions from light
  passenger and commercial vehicles (euro 6)(1),''
\newblock {\em Off. J. Eur. Union, L: Legis.}, vol. 55, pp. 16--24, 2012.

\bibitem{localization_survey}
H.~{Liu}, H.~{Darabi}, P.~{Banerjee}, and J.~{Liu},
\newblock ``{Survey of Wireless Indoor Positioning Techniques and Systems},''
\newblock {\em IEEE Transactions on Systems, Man, and Cybernetics, Part C
  (Applications and Reviews)}, vol. 37, no. 6, pp. 1067--1080, 2007.

\bibitem{indoor_cell_level_localization_based_on_RSSI_classification}
K.~{Lee} and L.~{Lampe},
\newblock ``Indoor cell-level localization based on {RSSI} classification,''
\newblock in {\em 2011 24th Canadian Conference on Electrical and Computer
  Engineering(CCECE)}, 2011, pp. 000021--000026.

\bibitem{RSSI_based_classification_for_indoor}
S.~{Mahfouz}, P.~{Nader}, and P.~E. {Abi-Char},
\newblock ``{RSSI}-based classification for indoor localization in wireless
  sensor networks,''
\newblock in {\em 2020 IEEE International Conference on Informatics, IoT, and
  Enabling Technologies (ICIoT)}, 2020, pp. 323--328.

\bibitem{life_cycle_ephesos}
H.~{Bernhard}, A.~{Springer}, A.~{Berger}, and P.~{Priller},
\newblock ``Life cycle of wireless sensor nodes in industrial environments,''
\newblock in {\em 2017 IEEE 13th International Workshop on Factory
  Communication Systems (WFCS)}, 2017, pp. 1--9.

\bibitem{SAL-RB_Dataset}
J.~Karoliny, T.~Blazek, F.~Ademaj, and H.~Bernhard,
\newblock ``{SAL-Autarkic-Localization-RSSI-BLE-Dataset: SAL- RB-Dataset},''
  Distributed by Zenodo {https://doi.org/10.5281/zenodo.4073072}, Oct. 2020.

\bibitem{scikit-learn}
F.~Pedregosa, G.~Varoquaux, A.~Gramfort, V.~Michel, B.~Thirion, O.~Grisel,
  M.~Blondel, P.~Prettenhofer, R.~Weiss, V.~Dubourg, J.~Vanderplas, A.~Passos,
  D.~Cournapeau, M.~Brucher, M.~Perrot, and E.~Duchesnay,
\newblock ``Scikit-learn: Machine learning in {P}ython,''
\newblock {\em Journal of Machine Learning Research}, vol. 12, pp. 2825--2830,
  2011.

\bibitem{rappaport1996wireless}
T.~S. Rappaport et~al.,
\newblock {\em Wireless communications: principles and practice}, vol.~2,
\newblock prentice hall PTR New Jersey, 1996.

\bibitem{TimeSeriesClassification_HiddenMarkov}
B.~{Esmael}, A.~{Arnaout}, R.~K. {Fruhwirth}, and G.~{Thonhauser},
\newblock ``Improving time series classification using hidden markov models,''
\newblock in {\em 2012 12th International Conference on Hybrid Intelligent
  Systems (HIS)}, 2012, pp. 502--507.

\end{thebibliography}

\end{document}